\documentclass{kluwer}
\usepackage[dvips]{graphicx}
\begin{document}
\begin{article}
\begin{opening}
\title{SiO Observations of the NGC 1333\\
 IRAS 2A Protostellar Jet}            

\author{F.P. \surname{Wilkin}\email{f.wilkin@astrosmo.unam.mx}\thanks{Partially supported by NSF International Researchers Fellows Program}}
\institute{Instituto de Astronom\'{\i}a, UNAM Morelia, M\'exico}
\author{H.-R. \surname{Chen}} 
\institute{Astronomy Department, University of California, Berkeley, USA}   
\author{L.M. \surname{Chernin}}
\institute{2694 Barkley Ave, Santa Clara, CA, USA}
\author{R.L. \surname{Plambeck}} 
\institute{Astronomy Department, University of California, Berkeley, USA}

\runningtitle{IRAS 2A Protostellar Jet}
\runningauthor{Wilkin e.a.}

\begin{abstract} 
We present 86.8 GHz SiO J=2-1 observations of the 
molecular outflow from the protostellar source NGC 1333 IRAS 2A. 
Silicon monoxide is a sensitive tracer of shocks in molecular gas, 
and the emission clearly traces terminal bowshocks at the ends of the outflow 
lobes, as well as emission knots and a source coincident with the protostar. 
Our three-field mosaic from the BIMA interferometer covers 
the entire outflow length, including both lobes.  The most prominent 
emission feature, the eastern bowshock, shows SiO at the leading edge, 
clearly displaced from the maximum  seen in CO. This is in contrast to
recent observations by Chandler \& Richer (2001) of  the HH 211 jet, 
where SiO traced primarily internal working surfaces and was less extended 
than the CO.  Our results are consistent with prompt entrainment at the
leading edge of the IRAS 2A bowshocks. 
\end{abstract}


\end{opening}
\section{Summary}
The class 0 source IRAS 2A drives an unusual, highly collimated
 jet-like outflow first studied in CO (Sandell {\it et al}.~1994). Such a
young, symmetric, and focused outflow provides an excellent 
opportunity to observe the expected bow shock of a jet interacting
with the molecular ambient gas.  We used the BIMA 
interferometer to map this outflow in SiO in order to trace shocks
within the outflow. Our initial three-field, C-array  (10''$\times 7''$ beam)
mosaic covers the positions of IRAS 2A and the east and west bow shocks. 
In addition to this outflow of roughly E-W
orientation, a larger, less collimated bipolar outflow originates from
nearly the same position (Liseau {\it et al.}~1988), 
and is possibly associated with the nearby
source IRAS 2C (Knee \& Sandell 2000; Sandell \& Knee 2001). 
Because the position of IRAS 2C agrees poorly with the symmetry axes of the
flows seen in CO, the possibility exists
that the two perpendicular outflows seen in CO might actually represent
a single, much wider, limb-brightened  bipolar outflow (see Figure~1). 
\begin{figure}[h]
\begin{center}
\centerline{\includegraphics[width=30pc,angle=270]{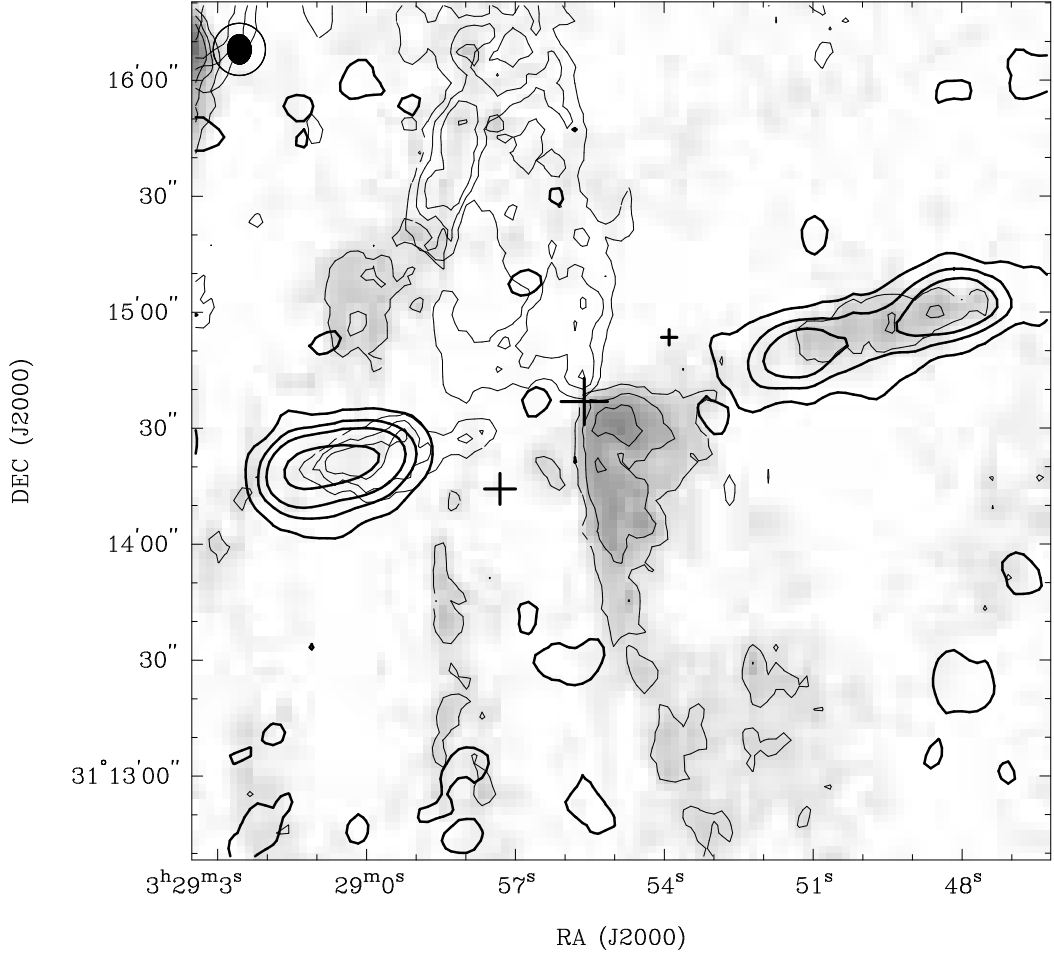}}
\caption{Mosaic of the IRAS 2A region. Thin contors show 
12CO J=2-1 taken from the BIMA observations of 
Engargiola \& Plambeck 1999. The greyscale indicates blue-shifted CO.
Thick contors show SiO J=2-1 emission. Crosses of decreasing size 
give the positions  of IRAS 2A, 2B, and 2C.}
\label{fig:map}
\end{center}
\end{figure}
Our recent observations with BIMA in the D-array (13.5'' beam) 
detect SiO in the E-W outflow but not the N-S, strongly suggesting that
these are two separate outflows.   
Nonetheless, it still appears likely that the driving source of the 
N-S flow remains to be discovered, and further searches for binarity
of IRAS 2A are suggested.  In Figures~2
and 3 we present the higher resolution wideband and channel maps of SiO
in the E-W outflow.
The much brighter, redshifted, eastern bow shock shows a wider range 
of velocities, and is associated with a dense clump seen in CS by
Sandell {\it et al.}, and whose V-shaped geometry resolved by Blake (1996)
suggests a bow shock driven into a steep density gradient. 
Presumably the eastern bow shock is much brighter
because the driving jet is interacting with this dense clump. 
The actual emission peaks in both the east and west lobes are unresolved
in our maps, and higher resolution observations are needed to confirm
the  expected bow shock geometry at the ends of the outflow lobes. 
\begin{figure}[h]
\begin{center}
\centerline{\includegraphics[width=30pc,angle=270]{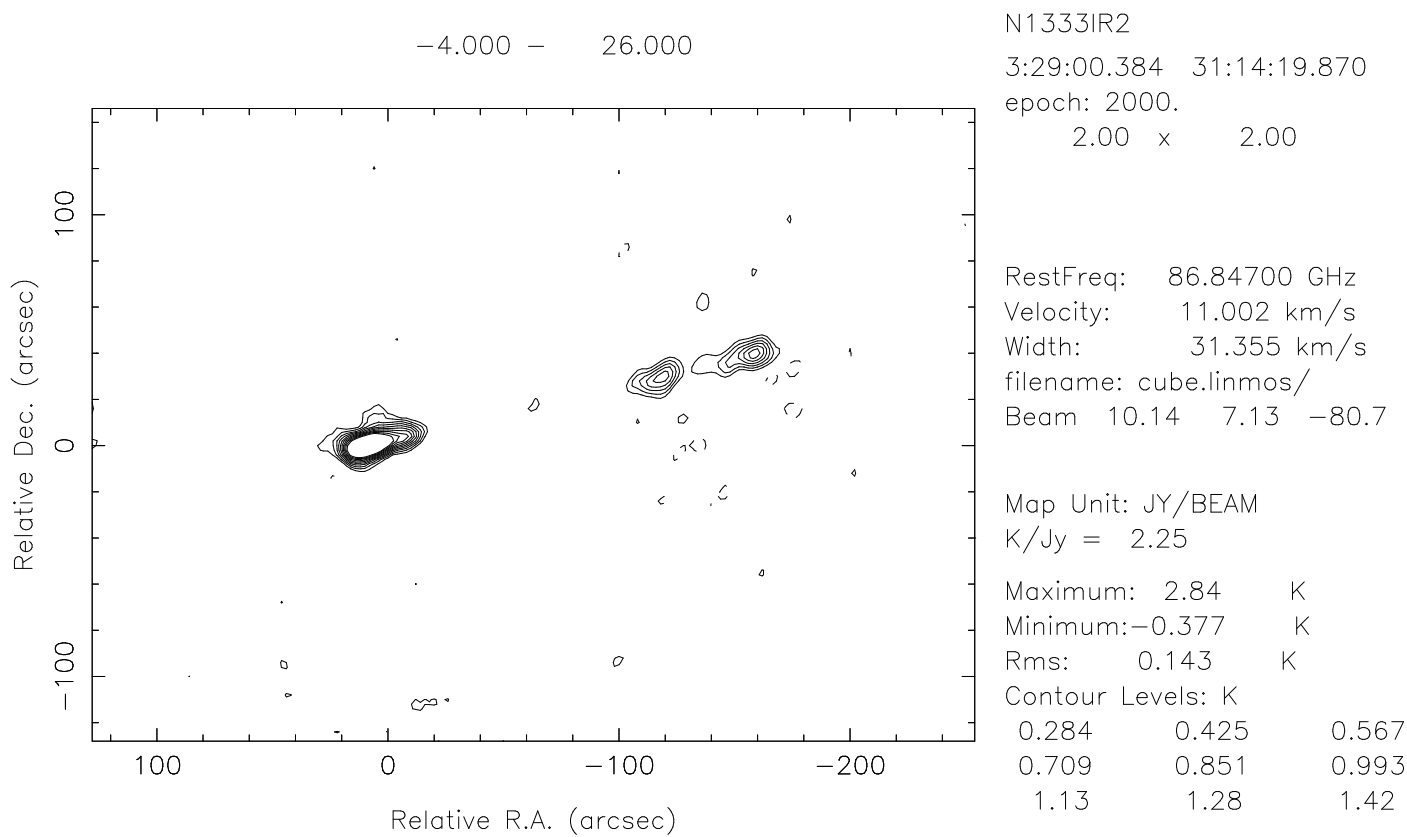}}
\caption{3-Field mosaic of IRAS 2A outflow in SiO j=2-1. In this map,
positions are relative to the bright, eastern emission peak. 
The submm point source locations are shown in Figures~1 and 3.}
\label{fig:map}
\end{center}
\end{figure}
\begin{figure}[h]
\begin{center}
\centerline{\includegraphics[width=30pc,angle=270]{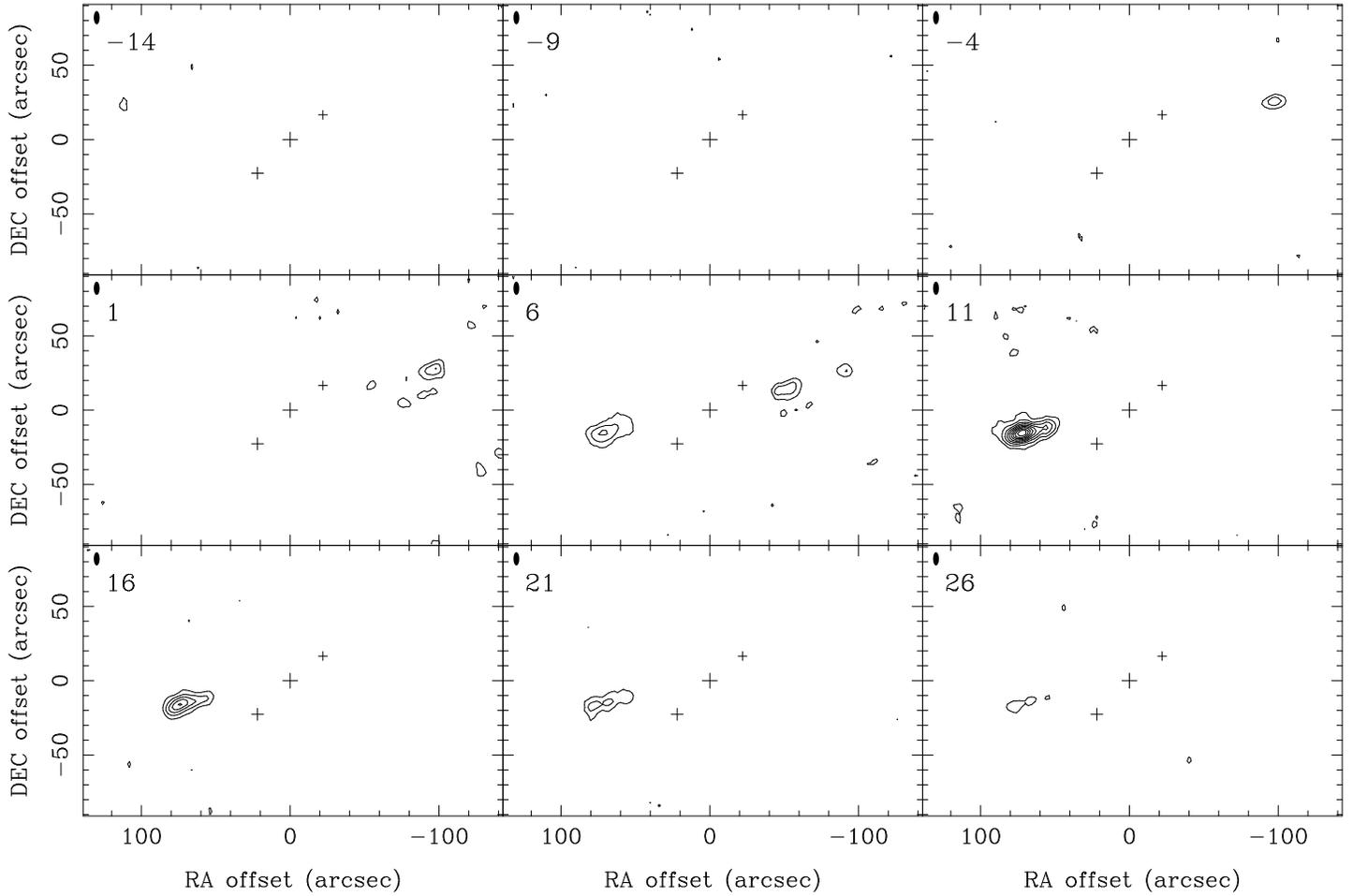}}
\caption{Channel maps at $5\,{\rm km}/{\rm s}$ intervals. The large, medium  
and small crosses indicate the positions of IRAS 2A, 2B and 2C, respectively.}
\label{fig:channels}
\end{center}
\end{figure}

\end{article}

\begin{thebibliography}{99}
\small
\itemsep 0pt
\bibitem[Blake(1996)]{blake} Blake, G.~A.\ 1996, IAU 
Symp.~178: Molecules in Astrophysics: Probes \& Processes, 178, 31 
\bibitem[Chandler \& Richer(2001)]{cr} 
Chandler, C.J. \& Richer, J.S. 2001, ApJ 555, 139
\bibitem[Engargiola \& Plambeck(1999)]{1999pcim.conf..291E} Engargiola, 
G.~\& Plambeck, R.~L.\ 1999, The Physics and Chemistry of the Interstellar 
Medium, Eds.: V.~Ossenkopf, J.~Stutzki, and G.~Winnewisser
\bibitem[Liseau (1988)]{liseau} 
Liseau, R., Sandell, G., Knee, L.B.G. 1988, A\&A 192, 153
\bibitem[Sandell et al.~(1994)]{sandell1994} 
Sandell, G., Knee, L.B.G., Aspin, C., Robson, I.E., \& Russell, A.P.G. 1994,
1994, A\&A 285, L1
\bibitem[Sandell \& Knee (2001)]{sn2001} 
Sandell, G., Knee, L.B.G. 2001, ApJ 546, L49
\end{thebibliography}
\end{document}